\documentclass[12pt]{article}
\usepackage{amssymb}
\usepackage{amsfonts}
\usepackage{amsmath}
\usepackage{latexsym}
\usepackage[T1]{fontenc}
\usepackage[utf8]{inputenc}
\begin{document}

\title{On a certain inequality for the sum of  norms and  reverse uncertainty relations}

\author{
Krzysztof  Urbanowski\footnote{e--mail:  K.Urbanowski@if.uz.zgora.pl}\;\,\footnote{Corresponding author}\\
Institute of Physics,
University of Zielona G\'{o}ra,  \\
ul. Prof. Z. Szafrana 4a,
65-516 Zielona G\'{o}ra,
Poland}
%\\
%\hfill\\
%e--mail:  K.Urbanowski@if.uz.zgora.pl}

\maketitle

\begin{abstract}
 We prove a simple inequality  for a sum of squares of  norms of two vectors in an inner product space. Next, using this inequality we derive the so--called "reverse uncertainty relation" and analyze its properties.
\end{abstract}

\noindent
Keywords: {\em
Inequality for  a sum of norms,   Reverse uncertainty relation.}
%\\
%Mathematics Subject Classification:  {\em Primary:} 46C05. {\em Secondary:} 	81Pxx,  81P15.\\

\section{Introduction}

In a linear spaces $ X$  with the inner product, one can prove many inequalities satisfied by vectors belonging to these spaces. They have a number of important applications not only in mathematics, but also in mathematical physics, and in particular in quantum mechanics. An example here is the Schwartz inequality used in the derivation of Heisenberg's quantum uncertainty principle. Other inequalities are used to derive  the so--called "sum uncertainty relations" (see, e. g. \cite{Pat,Mac}).
In some applications it is important to know the upper bound on the sum of norms. Such a bound can be found using, for example, the Dunkl--Wiliams inequality \cite{Dun}.
An example of this is its use in mathematical physics to derive the so--called "reverse uncertainty relation" \cite{Deb,Xiao3,Xiao4,Pri,Pri1,Pri2}.
%%%%%%%%%%%%
The reverse uncertainty relation was found as a result of the analysis of the so--called "stronger uncertainty relation" for the sum of variances $(\Delta_{\phi} A)^{2}$ and $(\Delta_{\phi} B)^{2}$,
\begin{equation}
(\Delta_{\phi} A)^{2}\,+\,(\Delta_{\phi} B)^{2}\;\geq \; \frac{1}{2}[\Delta_{\phi} (A + B)]^{2}. \label{Mac12}
\end{equation}
which has been  derived in \cite{Mac}. Here $A$ and $B$ represent non--commuting observables and $|\phi\rangle$ is the state of the quantum system under considerations.
The reverse uncertainty relation may open up new areas of research and applications in quantum physics, quantum metrology or quantum technologies.
Therefore, finding the simplest possible ways to write it in the language of observables and state vectors and making necessary computations easier seem to be extremely important.
Here we present a simple  inequality, (much simpler than that following from the Dunkl--Wiliams inequality), which seems to be a new,
 that can be used in the derivation of the above mentioned reverse uncertainty relation.

The paper is organized as follows: Section 2 contains a derivation of an inequality that can be used, among others, to derive the reverse uncertainty relation. Section 3 contains a new derivation  of the inverse uncertainty relation. In Sec. 4 you can find a discussion and concluding remarks.

\section{A certain simple inequality}
We start by proving the following theorem:\\
\hfill\\
\noindent
{\bf Theorem:}
Consider the linear vector space  $X$  with the inner product. Vectors,
$|\psi_{1}\rangle, |\psi_{2}\rangle$, belonging to such a space satisfy
the inequality \cite{X}
\begin{align}
\left\|\,|\psi_{1}\rangle\right\|^{2}\, &+\,\left\|\,|\psi_{2}\rangle\right\|^{2}  \leq  \left\|\,|\psi_{1}\rangle\, - \,|\psi_{2}\rangle\right\|^{2}\,+\,2\left|\langle\psi_{1}|\psi_{2}\rangle \right| \label{in0} \\
 &\leq  \left\|\,|\psi_{1}\rangle\, - \,|\psi_{2}\rangle\right\|^{2}
 +2\left\|\,|\psi_{1}\rangle \right\|\;\left\|\,|\psi_{2}\rangle\right\|. \label{in1}
\end{align}
{\bf Proof:}\\
Let's use the identity
\begin{align}
\left\|\,|\psi_{1}\rangle\right\|^{2}\,+\,\left\|\,|\psi_{2}\rangle\right\|^{2} & =\,\left\|\,|\psi_{1}\rangle\, - \,|\psi_{2}\rangle\right\|^{2}\nonumber \\
&\;\;\;\;\;+\,2 \,\Re(\langle \psi_{1}|\psi_{2}\rangle), \label{id1}
\end{align}
(where $\Re\,(z)$ is the real part of a complex number $z$), and the property,
\begin{equation}
\Re(\langle \psi_{1}|\psi_{2}\rangle) \leq \left|\langle \psi_{1}|\psi_{2}\rangle\right|, \label{in2a}
\end{equation}
then the Cauchy--Schwartz inequality,
we obtain
\begin{equation}
\left|\langle \psi_{1}|\psi_{2}\rangle\right|\,\leq \left\|\,|\psi_{1}\rangle\right\|\;\left\|\,|\psi_{2}\rangle\right\|. \label{in2}
\end{equation}
Replacing $\Re(\langle \psi_{1}|\psi_{2}\rangle)$ in (\ref{id1}) by (\ref{in2a}) we get the inequality (\ref{in0}) and then using  (\ref{in2}) we get the result (\ref{in1}).
$\square$

\section{Applications: A reverse uncertainty relation}
The inequality (\ref{in1}) is simple and may be useful in some applications:
It seems that it
should be of interest to, among others,  physicists studying the so--called "reverse uncertainty relations", see e.g. \cite{Deb,Xiao3,Xiao4,Pri,Pri1,Pri2}.
 In these papers an upper bound for a sum of norms was applied to define the reverse uncertainty relation for the sum of variances and
 properties of such a relation  were analyzed.
The  upper bound for the relation (\ref{Mac12})  and having more complicated form then that resulting from (\ref{in0}), (\ref{in1})  was found in \cite{Deb,Xiao3,Xiao4} using the Dunkl--Williams inequality \cite{Dun}.
Now let us try to derive the "reverse uncertainty relation" using inequalities (\ref{in0}) and (\ref{in1}).

%%%%%%%%%%%%%%%%%%
 In a general case,
   the variance $(\Delta_{\phi} F)^{2}$ of an observable $F$, when the quantum system is in the state $|\phi\rangle$, is defined as follows
\begin{equation}
(\Delta_{\phi} F)^{2} = \| \delta_{\phi} F|\phi\rangle\|^{2}, \label{d2F}
\end{equation}
where
$\delta_{\phi} F = F - \langle F\rangle_{\phi}\,\mathbb{I} $, and $\langle F\rangle_{\phi} \stackrel{\rm def}{=} \langle \phi|F|\phi\rangle$ is the  expected value of an observable $F$ in a system that is in the state $|\phi\rangle$, (where   $|\phi\rangle \in {\cal H}$ is the normalized vector and ${\cal H}$ is the Hilbert space of states of the quantum system under considerations), provided that $|\langle\phi|F|\phi \rangle |< \infty$.
Equivalently:
 $$(\Delta_{\phi} F)^{2} \equiv \langle F^{2}\rangle_{\phi} - \langle F\rangle_{\phi}^{2}.$$
 The observable $F$ is represented by hermitian operator $F$ acting in ${\cal H}$. Here $\Delta_{\phi} F \geq 0$ is the standard deviation. Let us consider
two observables, $A$ and $B$, represented by non--commuting hermitian operators $A$ and $B$ acting in ${\cal H}$, such that $[A,B]$ exists and $|\phi\rangle \in {\cal D}(AB) \bigcap {\cal D}(BA)$, (${\cal D}({\cal O})$ denotes the domain of an operator $\cal O$ or of a product of operators). Let
\begin{equation}
|\psi_{1}\rangle = \delta_{\phi} A|\phi\rangle \;\;{\rm and}\;\; |\psi_{2}\rangle = \delta_{\phi} B|\phi\rangle.  \label{dA-dB}
\end{equation}
If to insert (\ref{dA-dB}) into (\ref{in0}) then we obtain
\begin{align}
\left\|\delta_{\phi} A|\phi\rangle  \right\|^{2} + \left\|\delta_{\phi} B|\phi\rangle  \right\|^{2} & \leq  \left\|\delta_{\phi} A|\phi\rangle\, - \,\delta_{\phi} B|\phi\rangle\right\|^{2}\nonumber \\
&+2\left|\langle \phi|\delta_{\phi} A\;\delta_{\phi} B|\phi\rangle\right|. \label{rev-01}
\end{align}
There is $(\delta_{\phi} A|\phi\rangle\, - \,\delta_{\phi} B|\phi\rangle) \equiv \delta_{\phi} (A- B)|\phi\rangle $.  This and the definition  (\ref{d2F}) means that the inequality (\ref{rev-01}) takes the following form,
\begin{equation}
(\Delta_{\phi} A)^{2}\,+\,(\Delta_{\phi} B)^{2}\;\leq \; [\Delta_{\phi} (A - B)]^{2}\,+\, 2\left| {\cal C}_{\phi}(A,B) \right|, \label{rev-012}
\end{equation}
where
\begin{equation}
{\cal C}_{\phi}(A,B) = \langle \phi|\delta_{\phi} A\;\delta_{\phi} B|\phi\rangle \equiv \langle AB\rangle_{\phi} - \langle A\rangle_{\phi}\,\langle B\rangle_{\phi}, \label{c}
\end{equation}
is a quantum version of the covariance (or, of the correlation function)
of the observables $A$ and $B$ in quantum state $|\phi\rangle$.  Here defining the correlation function ${\cal C}_{\phi}(A,B)$ we follow, e. g.  \cite{Poz,Khr}.
The inequality (\ref{rev-01}) is a simple variant of the  "reverse uncertainty relation".

Another simple variant of the "reverse uncertainty relation" can be obtained using
(\ref{in1}).
Namely, applying the method used to derive the inequality  (\ref{rev-01}) to  (\ref{in1})
 and keeping in mind all steps leading to (\ref{rev-01}) we obtain that,
\begin{equation}
(\Delta_{\phi} A)^{2}\,+\,(\Delta_{\phi} B)^{2} \;\leq \;[\Delta_{\phi} (A - B)]^{2} \,+\,2\,\Delta_{\phi} A\cdot \Delta_{\phi} B, \label{rev-2}
\end{equation}
which is  another, less restrictive,  variant of the  "reverse uncertainty relation".

\section{Final remarks}

The Dunkl--Williams inequality for vectors $|\psi_{1}\rangle, |\psi_{2}\rangle$ from a real or complex inner product space has the following form \cite{Dun,Cer},
\begin{align}
\left\|\, |\psi_{1}\rangle - |\psi_{2}\rangle \right\| & \geq \frac{1}{2} \left( \left\|\,|\psi_{1}\rangle\right\| + \left\|\,|\psi_{2}\rangle\right\|\right) \times  \nonumber \\
 &\;\;\;\;\;\times \left\| \frac{|\psi_{1}\rangle}{\left\| \,|\psi_{1}\rangle\right\|} -  \frac{|\psi_{2}\rangle}{\left\| \,|\psi_{2}\rangle\right\|} \right\|, \label{Dun}
\end{align}
where the condition that  $|\psi_{1}\rangle, |\psi_{2}\rangle$ are nonzero vectors must be satisfied \cite{Dun,Cer}. This inequality was used in \cite{Deb} to find the "reverse uncertainty relation".
Indeed, replacing   $|\psi_{1}\rangle, |\psi_{2}\rangle$ in (\ref{Dun})  by (\ref{dA-dB}) and using the definition (\ref{d2F}) after some algebra Mondal and co--authors of  \cite{Deb} obtained that
\begin{equation}
(\Delta_{\phi} A)^{2}\,+\,(\Delta_{\phi} B)^{2} \;\leq \;2 \frac{[\Delta_{\phi} (A - B)]^{2} }{\left[1- \frac{cov_{\phi}(A,B)}{\Delta_{\phi} A\cdot \Delta_{\phi} B}\right]}\,-\,2\,\Delta_{\phi} A\cdot \Delta_{\phi} B, \label{Deb}
\end{equation}
where $ cov_{\phi}(A,B) = \Re\,[{\cal C}_{\phi}(A,B)]$. The inequality (\ref{Deb}) is the {\em reverse uncertainty relation} derived in \cite{Deb}.

As can be seen from the inequality (\ref{Deb}) this reverse uncertainty relation has rather complicated form and
is undefined if $|\phi\rangle$  is an eigenvector of $ A$ (or of $B$).
This is because then $|\psi_{1}\rangle = \delta_{\phi}A|\phi\rangle = 0$ (or  $|\psi_{2}\rangle = \delta_{\phi}B|\phi\rangle = 0$) and then the inequality (\ref{Dun}) does not hold.
%%%%%%%%%%%
%%%%%%%%%%%
These kinds of weaknesses are absent in inequalities (\ref{rev-01}) and (\ref{rev-2}).
Although inequalities (\ref{rev-01}) and (\ref{rev-2}) do not provide any useful information about the upper bound for the sum of two variances, $(\Delta_{\phi} A)^{2}\,+\,(\Delta_{\phi} B)^{2}$, if $|\phi\rangle$ is an eigenvector of $A$ (or $B$), but even in such a case the left and right sides of these inequalities are finite and well--defined, which cannot be said about inequality (\ref{Deb}).
%%%%%%%%%%%%%%%%%%%
Another weak point of the inequality (\ref{Deb})
can be found analyzing the following quantity
%is the case when
\begin{equation}
\frac{cov_{\phi}(A,B)}{\Delta_{\phi} A\cdot \Delta_{\phi} B} \stackrel{\rm def}{ =}q. \label{Deb1}
\end{equation}
It is obvious that $ q \leq 1$.
Writing $q$ as $q = 1 - \varepsilon$, where $\varepsilon \leq 0$ and $ \varepsilon \ll 1$, we see that the denominator in the first term of the right--hand side of inequality (\ref{Deb}) is simply equal to $(1 - q) \equiv \varepsilon $.
This means that the first term of the right--hand side of
this inequality
tends to infinity, when $\varepsilon \to 0$.
%%%
Then the upper bound on the sum of variances in inequality (\ref{Deb}) is equal to infinity and does not allow drawing any conclusions about the upper bound on the sum of variances allowed in the quantum system under study.
A similar effect occurs when $\varepsilon $ is very small and close to zero, $\varepsilon  \simeq 0$. This causes, similarly to the limiting case of $\varepsilon \to 0$, the upper bound on the sum of variances to become extremely large, making it impossible to draw any useful conclusions about it.
On the other hand, inequalities (\ref{rev-012}) and (\ref{rev-2}) are insensitive to the effects generated by properties of the quantity (\ref{Deb1}).
%%%%%%%%%%%%%%%%%%%%%%%%%
Moreover, inequalities (\ref{rev-012}) and (\ref{rev-2}) seem to be simpler in applications than inequality
(\ref{Deb}).

Reverse uncertainty relations,  (\ref{rev-012}),  (\ref{rev-2}) and  (\ref{Deb}) have another non--obvious property that is worth mentioning.
Namely, if the system is in such a state $|\phi\rangle$  that $|\psi_{1}\rangle = \delta_{\phi} A|\phi\rangle \;\perp \; |\psi_{2}\rangle = \delta_{\phi} B|\phi\rangle $, and simultaneously $\Delta_{\phi}A > 0$ and $\Delta_{\phi}B > 0$, then it is not possible to obtain any useful information about the upper bound for the sum of variances from these relations. Indeed, in this case ${\cal C}_{\phi}(A,B) = 0$,
and thus
$\Re\,[{\cal C}_{\phi}(A,B)] = cor_{\phi}(A,B) = 0$, and $[\Delta_{\phi}(A \pm B)]^{2} \equiv (\Delta_{\phi}A)^{2} + (\Delta_{\phi}B)^{2}$. The first observation is that in such a case observables $A$ and $B$ are uncorelated in this state.
Further observations are that in the situation under consideration the inequality (\ref{rev-012}) takes the form
$(\Delta_{\phi} A)^{2}\,+\,(\Delta_{\phi} B)^{2}\;\leq \;(\Delta_{\phi} A)^{2}\,+\,(\Delta_{\phi} B)^{2}$, and the inequality (\ref{rev-2}) looks as follows,
$ 0 \leq \Delta_{\phi} A\cdot \Delta_{\phi} B$, and finally, inequality  (\ref{Deb}) takes the form: $ 0 \leq(\Delta_{\phi}A \,-\,\Delta_{\phi}B)^{2}$.
Neither of these results says anything about the upper bound for the sum  $(\Delta_{\phi} A)^{2}\,+\,(\Delta_{\phi} B)^{2}$.
So if observables $ A$  and $B$ are uncorrelated in state $|\phi\rangle$  then using only inequalities (\ref{rev-012}),  (\ref{rev-2}) and  (\ref{Deb}) nothing can be said about the upper bound on the sum of their variances.

To sum up: inequalities  (\ref{rev-012}),  (\ref{rev-2}) and
(\ref{Deb})
are worth further investigations.
Such studies both are extremely important because
the reverse uncertainty relation shows that in the case of quantum phenomena, the possible dispersion of values of physical quantities represented by non--commuting observables $A$ and $B$
always has an upper bound.
Strictly speaking, as it results from  (\ref{rev-012}),  (\ref{rev-2}) and
(\ref{Deb}), there is an upper bound for the sum of variances.
This complements the conclusion resulting from the Heisenberg--Robertson uncertainty relation (and from the uncertainty relation for the sum of variances \cite{Mac}) that the spread of these values
cannot be arbitrarily small.

\section*{Acknowledgements }
The author would like to thank Janusz Matkowski for the inspiring discussion.
This work was supported by
the program of the Polish Ministry
of Science and Higher Education under the name "Regional
Initiative of Excellence", Project No. RID/SP/0050/2024/1.

\section*{Conflict of Interest}
The author declares no conflict of interest.

\section*{Competing Interest}
The author declares that there is not any personal, academic interest, or any other factors that may be perceived to influence the objectivity, integrity or value of the study.

\section*{Data Availability}
This manuscript has no
associated data, or the data will not be deposited. [Author's
comment: This is a theoretical work and analytical calculations are made. Therefore, no data are required].


\begin{thebibliography}{3}
\bibitem{X}
Functional analysis textbooks usually use the following notation for vectors from a linear vector space $X$: $\psi, \phi \in X$ instead of Dirac's notation $|\psi\rangle, |\phi\rangle \in X$.
Similarly, the inner product in such textbooks is commonly denoted as $(\psi,\phi)$ or $(\psi|\phi)$ instead of the common notation in theoretical physics $\langle\psi|\phi\rangle$. Generally, in this paper  $(\psi,\phi) \equiv \langle\psi|\phi\rangle$.
\end{thebibliography}
\end{document}